\documentclass{aa}

\usepackage{epsf}

\usepackage{times}
\usepackage{mathptm}

\newcommand{\rbf}[1]{#1}

\begin{document}
\thesaurus{03                          
              ( 11.17.3,               
                11.16.1,               
                11.19.1,               
                12.03.3,               
                04.19.1                
               )}

\title{The Hamburg/ESO survey for bright QSOs. III.
       A large flux-limited sample of QSOs
       \thanks{Based on observations made at the European Southern Observatory, 
               La Silla, Chile}
       }

\titlerunning{The Hamburg/ESO survey for bright QSOs. III.}

\author{L.~Wisotzki \and N.~Christlieb \and N.~Bade \and V.~Beckmann 
        \and T. K\"ohler \and C.~Vanelle \and D.~Reimers}

\institute{Hamburger Sternwarte, Gojenbergsweg 112, D-21029 
           Hamburg, Germany,\protect\\
           e-mail: {\tt lwisotzki@hs.uni-hamburg.de}
           }
          
\offprints{L.~Wisotzki}
\date{Received ; accepted} 
\maketitle

\markboth{L. Wisotzki et al.: The Hamburg/ESO survey. III.}{}

\begin{abstract}
We present a new sample of 415 bright QSOs and Seyfert~1 nuclei 
drawn from the Hamburg/ESO survey (HES). The sample is 
spectroscopically 99\,\% complete and well-defined 
in terms of flux and redshift limits. 
Optical magnitudes are in the interval $13\la B_J\la 17.5$, 
redshifts range within $0< z < 3.2$.
More than 50\,\% of the objects in the sample are new discoveries.
We describe the selection techniques and discuss
sample completeness and potential selection effects.
There is no evidence for redshift-dependent variations of 
completeness; in particular, low-redshift QSOs --
notoriously missed by other optical surveys -- 
are abundant in this sample, since no discrimination against 
extended sources is imposed.
For the same reason, the HES is not biased against QSOs 
multiply imaged due to gravitational lensing.
The sample forms the largest homogeneous set of bright QSOs
currently in existence, useful for a variety of statistical studies.
We have redetermined the bright part of the optical
quasar number-magnitude relation. We confirm that the Palomar-Green
survey is significantly incomplete, but that its degree of
incompleteness has recently been overestimated. 

\keywords{
          Quasars: general -- Galaxies: photometry -- 
          Galaxies: Seyfert -- Cosmology: observations -- Surveys
          }

\end{abstract}

\section{Introduction}

While the total number of known quasars is growing rapidly,
most of these are too faint for detailed investigations. 
Existing catalogues are mostly inhomogeneous in composition 
and not representative, featuring disproportionately 
high fractions of radio-loud, X-ray- and infrared-selected QSOs
especially among the brightest known QSOs.
To homogeneously sample the bright part of the quasar population 
in the optical regime, substantial fractions of the extragalactic sky
have to be covered with efficient surveying techniques.
The only presently available such catalogue, the Palomar-Green
Bright Quasar Sample (BQS; Schmidt \& Green \cite{schm+gree:83:QEBQS}), 
is now known to be substantially incomplete
(Goldschmidt et al.\ \cite{goldschmidt*:92:SDBQ}; K\"ohler et al.\
\cite{koehler*:97:LQLF}). A similar project in the southern
hemisphere, the Edinburgh/Cape Survey 
(ECS; Stobie et al.\ \cite{stobie*:97:EC1}), has just produced
first output, and the completeness of this survey remains to be
assessed. Moreover, both BQS and ECS discriminate against
low-$z$ QSOs with extended host galaxies, and the photometric 
UV excess selection technique 
confines both surveys to $z<2.2$ QSOs.

In 1990 we have started the `Hamburg/ESO survey' (HES),
a new wide-angle survey for bright QSOs and Seyferts, as 
an ESO key programme. The survey is
based on digitised objective-prism photographs 
taken with the ESO Schmidt telescope, covering essentially the entire
southern extragalactic sky. QSO candidates are selected 
with largely automated procedures, minimising human interaction
and establishing well-defined input samples for follow-up slit
spectroscopy.
A description of initial design and first results of the HES
was given by Wisotzki et al.\ (\cite{wisotzki*:96:HES1}; 
hereafter Paper~1), where also 
the driving science objectives behind the survey have been listed.
A list of 160 newly discovered QSOs and Seyfert~1 galaxies
was published by Reimers et al.\ (\cite{reimers*:96:HES2}; 
hereafter Paper~2).
The first flux-limited sample of 55 QSOs, 
distributed over an effective area of 611\,deg$^2$, 
was presented and analysed by K\"ohler et al.\
(\cite{koehler*:97:LQLF}), to measure the surface density of bright
quasars and to estimate the combined 
local luminosity function of quasars and Seyfert~1 nuclei.

In this paper we present a major expansion of our earlier work.
The survey area has been sixfolded, and the QSO selection techniques
have evolved significantly. We have constructed a new flux-limited
sample of 415 optically bright QSOs that will be useful for a wide
variety of statistical investigations. We describe the selection
techniques used to build the sample and discuss its completeness,
in comparison with other surveys as well as in terms of expected
redshift-dependent selection effects.
In providing these details, this paper serves also as a reference
for future publications of HES-selected QSO samples.
The paper closes with a reassessment of the surface density of 
bright QSOs. In a companion paper (Wisotzki \cite{wisotzki:99:BQLF})
we study the impact of this new sample on the bright end of 
the QSO luminosity function and its evolution. Further work is
in progress to follow up on several issues like host galaxy
characteristics, radio properties, or the incidence of gravitational 
lensing events.

\begin{figure}[tb]
\epsfxsize=\hsize
\epsfclipon
\epsfbox[90 190 525 602]{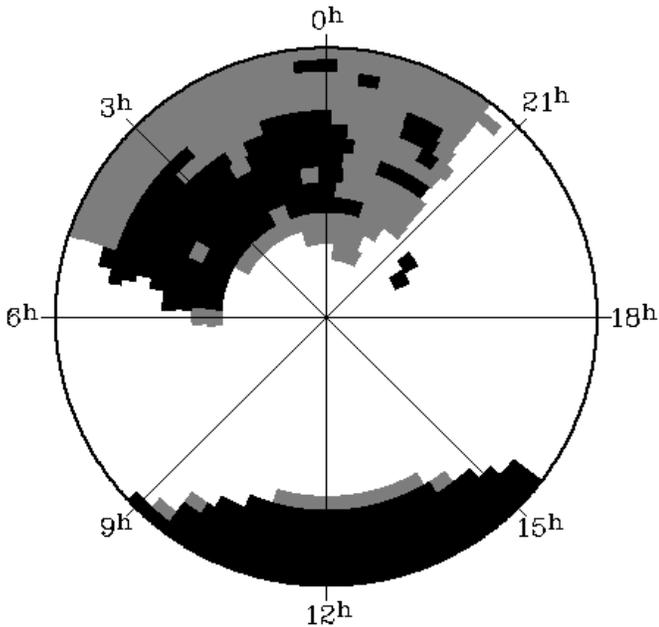} 
\caption[]{Distribution of HES survey fields in the southern hemisphere,
in area-conserving azimuthal projection. The South pole is in the centre, 
and meridians of constant right ascension are marked. 
The black areas correspond to the 207 fields forming the basis for the 
flux-limited sample, while greyshaded zones indicate further HES
fields where survey work is not yet completed.
Fields left white are in or too close to the Milky Way.}
\label{fig:area}
\end{figure}

\section{The Hamburg/ESO survey  \label{sec:HES}}

\subsection{Survey area and plate material}

Following the installation of the Hamburg/ESO survey,
444 objective prism plates were taken between 1990 and 1998
at the 1\,m Schmidt telescope on La Silla. The survey area has been
defined chiefly by the need to avoid regions too close to the Milky Way,
apart from the initial requirement $\delta < +2\fdg 5$. The basic
condition was a column density of Galactic neutral hydrogen
in the field centre of $N_H < 10^{21}$\,cm$^{-2}$,
with the $N_H$ data taken from the collation of Dickey \& 
Lockman (\cite{dick+lock:90:H1}), available through the EXSAS
software package.
Some further fields were excluded because of too high stellar
surface densities, as the reduction of effective area due to 
overlapping spectra increases dramatically with this number.
There are 380 ESO fields meeting these criteria, covering 
a total area of almost $\sim$ 10\,000\,deg$^2$ in the sky. 
Fig.\ \ref{fig:area} shows the distribution of the
380 survey fields in the sky, of which 207 were fully processed
at the time of constructing the new sample.
This included plate digitisation, reduction, photometric
calibration, as well as spectroscopic follow-up observations of
quasar candidates to well-defined magnitude limits. 
These 207 fields are marked in black in Fig.\ \ref{fig:area}.

The photographic material used for the spectral plates was always
unfiltered Kodak IIIa-J, giving a spectral range of 
\mbox{$\sim 3200$--5400\,\AA} .
The ESO Schmidt objective prism has a reciprocal dispersion of 
450\,\AA/mm at H$\gamma$, allowing for a 
(seeing-limited) spectral resolution of objective prism spectra
as high as 10\,\AA\ at H$\gamma$.
While in the past such high spectral resolution has always been regarded as
not useful for quasar work because of the bright detection limits,
it is in fact an extremely well-suited instrumentation 
to perform an efficient wide-angle search for bright QSOs.

For each spectral plate, a corresponding direct photograph is required.
The SERC-J atlas of the southern sky provides a natural counterpart.
The atlas has already been fully digitised at the STScI in Baltimore
and is readily available as part 1 of `The Digitized Sky Survey' (DSS)%
 \footnote{Based on photographic data obtained using The UK Schmidt Telescope.
 The UK Schmidt Telescope was operated by the Royal Observatory
 Edinburgh, with funding from the UK Science and Engineering Research
 Council, until 1988 June, and thereafter by the Anglo-Australian
 Observatory.  Original plate material is copyright (c) the Royal
 Observatory Edinburgh and the Anglo-Australian Observatory.  The
 plates were processed into the present compressed digital form with
 their permission.  The Digitized Sky Survey was produced at the Space
 Telescope Science Institute under US Government grant NAG W-2166.
 }
on CD-ROM. Since the nominal centres of spectral and direct
plates correspond and the UKST plates are much larger than the ESO plates,
there is a 100\,\% match of areas in virtually all fields,
even in the (quite common) cases in which the ESO prism plates were found 
to be decentered by several minutes of arc. The plate scale is nearly
the same for ESO Schmidt and UKST. A drawback of the DSS data is that 
the digitisation grid is rather coarse, allowing for a spatial resolution 
of $\sim 3''$ at best, much poorer than on the spectral plates. 
We found that many of the objects appearing extended in the DSS 
because of two merged stellar images
were in fact well separated in the digitised spectral data.

\begin{figure}[tb]
\epsfxsize=\hsize
\epsfclipon
\epsfbox[67 86 341 257]{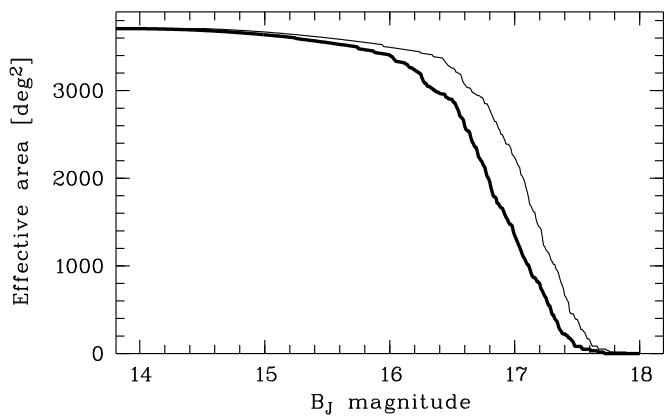}
\caption[]{Effective area of the surveyed region
as a function of $B_J$ magnitude (thin line: without correction
for Galactic extinction; thick line: with extinction correction).}
\label{fig:effarea}
\end{figure}

In a survey like this, it is not possible to utilize the entire
area of 27\,deg$^2$ formally subtended by the photographic plates for the 
selection of quasar candidates. Several effects contribute to a reduction 
of usable area:
\begin{itemize}
\item Decentred prism plates leave small gaps between adjacent fields,
as the ESO Schmidt plates are too small to always ensure sufficient
plate-to-plate overlap.
\item Bright stars, large galaxies, and a few globular clusters 
render certain regions of each plate unprocessable.
The same effect occurs if large-scale plate flaws are present.
\item Two nearby sources will appear as one single, merged object on the DSS direct
image, and the fainter of these will be lost for the input catalogue.
\item Most important is the fact that all slitless spectroscopic surveys 
suffer from losses due to overlapping spectra. These losses 
are particularly important for the unfiltered high-dispersion 
spectra of the ESO Schmidt. The loss factor depends strongly on
the stellar surface density and thus on Galactic latitude.
\end{itemize}
We have quantified all these effects,
incorporated them into `effective areas' $\Omega_{\mathrm{eff}}$.
As a grand average over all fields, 
the nominal area has to be reduced by roughly 25\,\% to obtain
$\Omega_{\mathrm{eff}}$; for individual fields the reduction factors
range between $<10$\,\% close to the South Galactic Pole and
$> 50$\,\% in the fields closest to the Milky Way.
Because the overlap rate depends on brightness, and because of
the field-to-field variations in the magnitude limits, 
the total effective survey area is a function of apparent magnitude.
For the 207 HES plates forming the basis for the sample discussed here,
the maximum effective area is 3700\,deg$^2$ for $B_J <14.5$,
decreasing gradually to zero around $B_J\simeq 17.7$ 
as shown in Fig.~\ref{fig:effarea}. 
A comprehensive listing of the properties of the 207 fields is given 
in Table \ref{tab:app:fields} of the appendix.

\subsection{Plate digitisation and data reduction}

All spectroscopic plates used for the HES were digitised using 
the Hamburg PDS 1010G microdensitometer, with the hardware upgrades as
described by Hagen et al.\ (\cite{hagen*:95:HQS1}). However,
the two-step digitisation procedure developed for the northern 
`Hamburg Quasar Survey', outlined in the same paper, has been
replaced for the HES in 1994/95 by full matrix scans of the entire
plates. Technical details of the digitisation and reduction procedure
will be given elsewhere; here we only comment briefly on a few important
properties that are relevant for the discussion below.
\begin{description}
\item[\normalfont\textit{Object detection}]
Source catalogues are constructed from the DSS direct images, 
with a magnitude limit of $B_J\simeq 21$.
\item[\normalfont\textit{Morphological classification}]
Most quasar surveys limit the selection of candidates to 
objects with `stellar' morphology. Within the HES, morphological 
classification into point and extended sources has been 
installed \emph{only} to optimise the following extraction of spectra.
\item[\normalfont\textit{Wavelength calibration}]
An astrometric transformation between direct and spectral plate
establishes for each dataset the precise location of objects and
provides wavelength zeropoints with an accuracy of typically 
$\sim 10\,\mu$m rms, or $\pm 5$\,\AA\ near H$\gamma$. 
\item[\normalfont\textit{Extraction of spectra}]
We have designed an optimal extraction algorithm of fitting
point-spread functions with pre-determined widths and positions 
to the 2--3 central pixels of the submatrix for each object,
leaving only the amplitudes as free parameters. This procedure 
minimises contamination from nearby sources, while at the same 
time the extraction windows (and subsequent photometry) are limited 
to apertures of the size of the seeing disk.
\item[\normalfont\textit{Detection limits}]
The formal detection limit is fixed at an internal magnitude 
corresponding to 2$\sigma_b$, where $\sigma_b$ is the rms 
background noise in the spectral plate. Typically, this corresponds 
to a $B_J$ magnitude of $\sim 18$--18.5.
The direct plate input catalogues reach much fainter,  
so the final candidate search is nearly independent of the 
object magnitudes on the direct plates. 

The bright limit of the survey is set by saturation of the
photographic spectra, when UV excess objects are no longer separable.
A special extraction scheme for partly saturated spectra allows to 
extend the bright limit up to $B_J \simeq 11$, yielding a
full dynamic range of the HES of $\sim$ seven magnitudes.
\end{description}

\section{Selection of quasar candidates \label{sec:cansel} }

While in most photographic surveys the selection of QSO candidates
was limited to \rbf{one criterion only}, the automated processing of 
multichannel data allows the user to apply several selection 
criteria in parallel, as first shown by Clowes et al.\ 
(\cite{clowes*:84:AQD}) and Hewett et al.\ (\cite{hewett*:85:PRS}).
For the HES we decided to \rbf{use} an even larger variety of descriptors
for the selection, including a certain redundancy with respect
to given spectral properties. Two sets of criteria are used,
`colours' and spectral feature detection.

\subsection{Colour selection} \label{sec:coloursel}

The first set of criteria mainly exploits
continuum properties and can be, to a certain extent, related to traditional 
photometric colours. The spectral energy distribution is parametrised using
the point of flux bisection (i.e.\ 50\,\% of the flux is located on \rbf{each side}),
or `half power point' (hpp). This concept is applied in three flavours:
\begin{description}
\item[\tt hpp1:] Bisecting point between 3200 and 4840\,\AA .
This number is well correlated with ($U-B$), cf.\ below.
\item[\tt hpp2:] Dito between 3950 and 5400\,\AA , approximately
 corresponding to $B-V$.
\item[\tt hpp3:] Dito between 3520 and 5050\,\AA . This is a formally 
redundant feature included to improve the detection rate at faint magnitudes.
\item[\tt qd] (quartile distance): The distance between the points
separating the first and the last 25\,\% of the flux. This feature
typically takes large values when the spectrum has composite nature, 
with a red `head' and a distinctly `blue' tail and has been tailored to
detect low-$z$ Seyfert galaxies.
\end{description}
This definition of `colours' has the advantage of being less
affected by noise than the simulated broad-band photometry used
during the first years of the HES (cf.\ Paper~1).
A similar but much simpler measure was already discussed
by Hewett et al.\ (\cite{hewett*:95:LBQS6}), who employed just 
a single half-power point bisecting the entire spectra.

\begin{figure}[tb]
\epsfxsize=\hsize
\epsfclipon
\epsfbox[64 86 341 257]{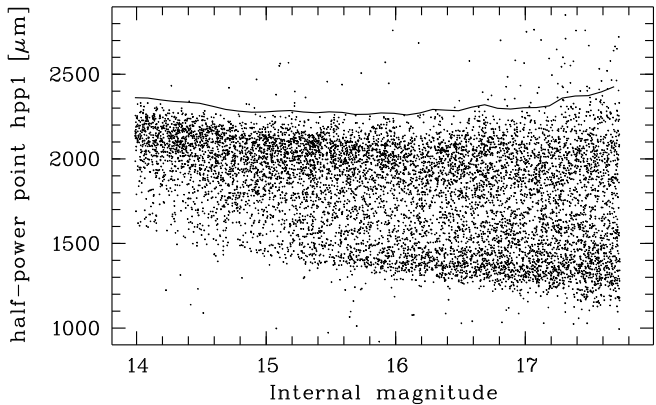} 
\caption[]{Selection of UV excess quasar candidates:
Distribution of `colour' \texttt{hpp1} as a function
of internal, uncalibrated magnitude. The solid line
separates the UV excess objects from the bulk of normal stars.}
\label{fig:coloursel}
\end{figure}

The selection of QSO candidates proceeds as follows: 
For a given dataset (scanned plate), all objects
are projected into two-dimensional arrays, with the internal 
PDS magnitude as one, and one of the `colours' as the other coordinate 
(cf.\ Fig.\ \ref{fig:coloursel}).
A cutoff-location algorithm then traces the well-defined
sharp blue edge of the distribution of Galactic stars,
approximately located at spectral types G--F. 
This yields a polygon adapted to the empirical distribution, 
independent of systematic variations from plate to plate, 
and also accounting for magnitude-dependent trends (such as 
emulsion nonlinearity) within a plate. 
Objects located on the blue side of the 
cutoff-tracing curve in each of these feature diagrams
are then marked as candidates.

\begin{figure}[tb]
\epsfxsize=\hsize
\epsfclipon
\epsfbox[64 86 341 257]{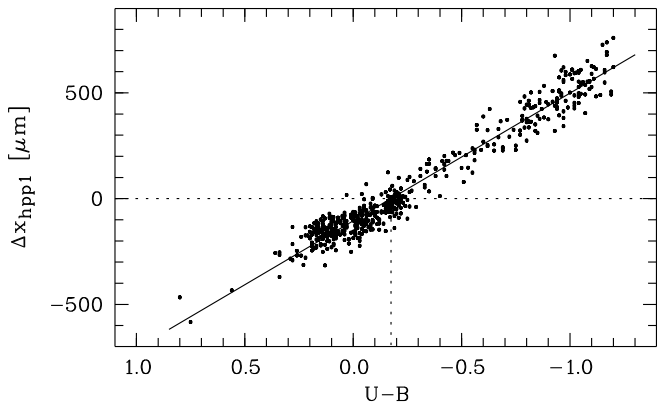} 
\caption[]{Comparison of UVX selection criterion with photoelectric
photometry: \texttt{hpp1} distance to selection cutoff vs.\
photoelectric ($U-B$).
Objects with $\Delta x_{\mathtt{hpp1}} > 0$ are selected
as UV excess candidates.}
\label{fig:uminbcal}
\end{figure}

To compare the performance of our colour selection scheme with standard
photometric selection, we have used published photolectric 
\emph{UBV} measurements of OB subdwarfs and metal-poor halo stars
found in our fields, taken from the Edinburgh/Cape UV excess survey 
(Kilkenny et al.\ \cite{kilkenny*:97:ECS2}) and the HK narrow-band prism
survey (Preston et al.\ \cite{preston*:91:HK}). Fig.\ \ref{fig:uminbcal}
shows the relative \texttt{hpp1} colours (distance to actual cutoff) 
of these stars plotted against ($U-B$). In this plot, objects above the
horizontal dotted line are selected as QSO candidates.
The correlation is excellent; it is nearly linear over the entire 
range, and the rms scatter is only 0.09~mag (a quadratic relation reduces
the scatter to 0.08\,mag). Notice that the plot was
compiled from several photographic plates; hence, the scatter is partly caused
by unavoidable plate-to-plate variations, and the true uncertainty of
converting \texttt{hpp1} colours into ($U-B$) \emph{within} one scan dataset 
will therefore be even smaller.
This astonishingly high accuracy of colour estimation can be explained 
partly by the fact that, in contrast to traditional photographic photometry, 
no external calibration is needed; but it also
demonstrates the quality and homogeneity of the ESO plate material.

The selection threshold shown in Fig.\ \ref{fig:uminbcal}
is equivalent to selecting objects with $(U-B) < -0.18$ (vertical dotted
line), considerably less restrictive than the conventional UV excess 
criterion $(U-B)\la -0.4$ imposed in most photometric surveys. Together
with the above uncertainty estimate this implies that of the objects 
with intrinsic $(U-B) < -0.5$, less than $\sim 1$\,\% will be missed
by the survey UVX selection criterion. 

We note in paranthesis that a similar relation 
exists between ($B-V$) and our second continuum `colour' \texttt{hpp2}. 
This colour has been included to widen the redshift range sensitive to 
colour selection schemes (cf.\ below).

\subsection{Feature detection}

According to the traditional approach, selecting quasars in sets of 
slitless spectra means looking for emission-line objects. Although
the majority of our objects are selected by colours, we have 
implemented such a search via a template matching algorithm.
This comprises cross-correlation of each continuum subtracted 
spectrum with Gaussian emission-line templates of various
widths, as well as with a differentiating `break locating' template of
zero total flux.

For each line detected by the matching algorithm, a signal-to-noise
value is computed as well as line flux and equivalent width;
all detections stronger than S/N\,=\,2.5 and with
$W_0 > 20\,\mu$m, or $\approx 10$\,\AA\ at H$\gamma$, are recorded.

A problem with these automatically detected features originates in
the high quality of the spectra themselves:
At the spectral resolution of the ESO Schmidt objective prism, 
common late-type stars show already very structured absorption line 
complexes, in particular in the near UV around and shortwards of 
the Ca\,{\sc ii} H+K doublet (which is clearly resolved). 
The automatically determined pseudo-continuum is systematically too low 
in such objects, causing many spurious detections of `emission lines' 
or `breaks' that are in fact just caused by absorption complexes. 
A selection only with respect to the S/N would therefore
yield thousands of false detections, and a more refined scheme
was required in order to make automated feature detection
viable. To accomplish this, all detections in a given scan are 
represented again as points in two-dimensional feature diagrams,
with the template S/N as one and the position in the spectrum
as the other coordinate. A cutoff-locating algorithm very similar to 
that employed in the colour selection provides a separation curve between
the tail of the stellar distribution and true outliers.
In consequence, the detection threshold is a nontrivial function of
wavelength: A typical selection threshold is S/N\,$>$\,5--6 for most
wavelengths, but increases to S/N\,$\ga$\,15 in narrow wavelength 
intervals where affected by common stellar absorption features. 
This makes the effiency of emission-line based
quasar selection also depending heavily on redshift $z$, 
with certain almost `blind spots' in redshift space. 
As we discuss below, this becomes relevant only at redshifts
$\ga 2.6$, while at all lower $z$ colour techniques are generally 
more efficient.

\subsection{Elimination of false candidates}

\begin{figure}[tb]
\epsfxsize=\hsize
\epsfclipon
\epsfbox[67 86 314 349]{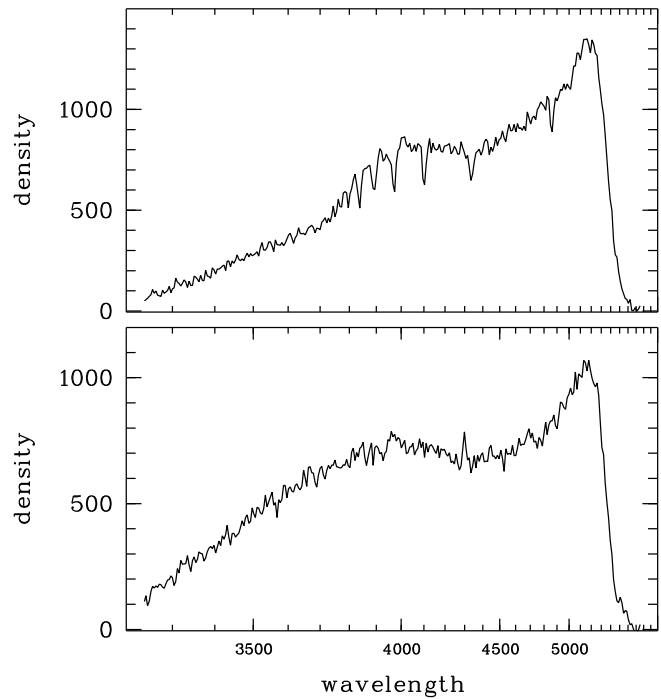} 
\caption[]{Example for digitised objective prism spectra, 
illustrating the potential of using high-dispersion slitless
data. The upper spectrum shows a subluminous blue star with 
narrow Balmer absorption lines, while the lower spectrum 
represents a low-redshift quasar ($z=0.15$), devoid of strong 
emission features except the weak [O\,{\sc ii}] line at 4300\,\AA .}
\label{fig:examples}
\end{figure}

After applying the primary selection criteria, the number of 
QSO candidates per plate is typically a few hundred per plate,
still considerably higher than the expected number of true quasars.
These automatically selected candidates are then subjected to an 
interactive check at the computer screen, 
where the extracted spectra are displayed together with the
corresponding pixel matrix images of direct and spectral plate.
It is generally straightforward to eliminate the dominant contaminations 
of false, i.e.\ obviously non-quasar candidates, 
from these candidate sets.
\begin{description}
\item[\normalfont\textit{Plate flaws and artefacts,}] also satellite
trails, ghost images etc.\ 
already contribute typically more than 50\,\% of the selected
candidates, in fact most of the feature detections.
By visual inspection of the direct or spectral plate images, such 
structures are easily identified.
\item[\normalfont\textit{Blue stars}] are always the main population 
in UV excess selected
samples. For the magnitudes sampled by the HES, the number ratio
between stars and quasars is typically $> 10:1$, further enhanced by the
broad range of selection criteria and the relaxed limits. However, the
large majority of these stars shows conspicuous Balmer absorption
lines easily discernible in the high-resolution spectral data,
as illustrated in Fig.\ \ref{fig:examples}, and 
are thus readily removed. For sufficient continuum signal-to-noise ratio
even the relatively weak-lined SdB and SdF stars, generally the most 
notorious contaminating population in quasar surveys, can be identified 
with high confidence from the objective-prism spectra alone.
\end{description}

\subsection{Follow-up spectroscopy}

Despite the good spectral resolution of the ESO Schmidt 
objective prism, slit spectra are generally superior in 
wavelength coverage, calibration accuracy, and also in 
signal-to-noise ratio.
A spectroscopic follow-up programme for high-grade QSO
candidates was started in November 1990 using the
1.52\,m, 2.2\,m, and 3.6\,m telescopes at ESO.
The spectra have typically $\sim 20\,$\AA\ resolution and
continuum signal-to-noise ratios of $\ga 15$, so that
unambiguous classifications and redshifts could be
assigned in most cases, without needing to obtain another exposure
(for several example spectra see Paper~2).
A number of spectra show no evidence of either emission
or absorption lines, despite a good S/N. 
While most of these objects are probably weak-lined white dwarfs,
it cannot be excluded that this group contains also a few BL~Lac
objects. These uncertainties will be probably resolved after a
cross-correlation of the HES with the ROSAT All-Sky Survey,
which is under way (cf.\ Bade et al.\ \cite{bade*:95:RASS}). 
At any rate, given the high S/N of most of the spectra 
it is unlikely that there are unrecognised quasars among these.
Objects already listed in the catalogue by 
V\'{e}ron-Cetty \& V\'{e}ron (\cite{vero+vero:96:AGN} and earlier
versions) were not generally observed, except 
when the catalogue entry contained no redshift, or
if the objective prism spectrum gave rise to suspicion 
that the catalogue redshift might be wrong.

As already mentioned in Paper~1, the HES follows a twofold 
selection strategy: Close to the plate detection limit, 
only \emph{bona fide} QSO candidates, in particular at high redshift, 
are followed up spectroscopically.
Operational completeness of candidate follow-up is attempted only for a 
brighter limiting magnitude. This `completeness limit' corresponds
to a minimum S/N of 5 in the spectra, approximately one magnitude
brighter than the detection  limit. In some fields, 
the follow-up had to be \rbf{stopped} at a yet brighter but still
well-defined limit.
Altogether, 587 objects were observed for the flux-limited sample,
with only two further sources ($<0.3$\,\%) 
still lacking spectroscopic identification.

An important measure for the efficiency of a quasar survey is the 
relative number of `failures' in the follow-up spectroscopy, indicative of
the ability to discriminate between quasars and stars. Note that 
this number as such is not yet an efficiency indicator, 
as it depends also strongly on the magnitude limit.
Brighter surveys inevitably show a higher fraction of stars
because of the very different number-magnitude relations of quasars
and stars. For example, the Palomar-Green survey for bright 
($B\la 16.2$) UV excess objects  (Green et al.\ (\cite{green*:86:PG}) 
contains a stellar fraction of $\sim 93\,\%$, 
while three magnitudes fainter the number ratio of UVX stars to quasars 
is about 1:1 (La Franca et al.\ \cite{lafranca*:92:SA94}). 
The stellar fraction for the HES colour-selected candidates is $\ga 90\,\%$
(see above);
it is fortunate that most of these objects are SdB stars or DA white 
dwarfs, since precisely for these classes the recognition of stellar 
absorption features in the objective prism spectra is extremely efficient.
These obviously stellar objects were not even included in the follow-up lists,
leading to a dramatically reduced stellar contamination among the quasar 
candidates to be observed: Only 189 out of the 587 
observed candidates for the `complete sample' turned out to be stars, 
most of these of rare or even highly peculiar nature (e.g., magnetic
white dwarfs, cf.\ Reimers et al.\ \cite{reimers*:96:FWD}).
This extremely high success rate of $\sim 2$:1 is unmatched for an
optically bright quasar survey, 
approaching the efficiencies obtained in X-ray and
radio follow-up identification programmes.
Note that this rate has been achieved \emph{without} sacrifices in
completeness, as it is entirely due to the enhanced ability of the
HES to eliminate the stellar contamination.

\begin{figure}[tb]
\epsfxsize=\hsize
\epsfclipon
\epsfbox[72 86 341 257]{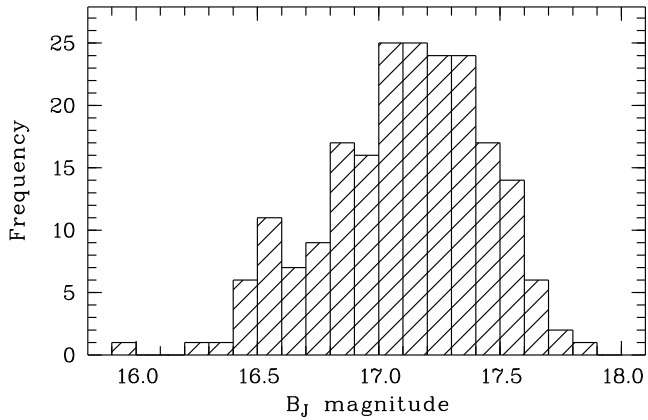} 
\caption[]{Distribution of actual $B_J$ 
magnitude limits for the 207 HES fields used for the complete sample.}
\label{fig:maglim}
\end{figure}

\subsection{Photometry}

In contrast to photometric surveys, QSO search on objective prism
plates does not \emph{require} prior photometric calibration, as all
quantities including colour information can be expressed consistently
in internal units. Photometric measurements of all 
individual sources as well as the determination of survey flux limits,
tied to a homogeneous external photometric system, become essential
only when statistical quantities such as surface densities and
luminosity functions are to be determined.
For a wide-angle survey such as the HES, the \rbf{main} difficulty to
overcome was the lack of suitable photometric standard stars in most
of the survey fields. Although since publication of the HST \emph{Guide
Star Photometric Catalogue} (GSPC; Lasker et al.\ \cite{lasker*:90:GSPC1})
there are at least a few standard stars known in each ESO
field, these stars generally do not cover the magnitude range of
interest ($B\la 15$ for the GSPC, while $B = 15$--18 is needed).
In an attempt to obtain moderately deep ($B\la 19$) sequences of \emph{all}
380 HES survey fields, we have launched a photometric calibration programme 
using the ESO/Dutch 90\,cm telescope.
A full description of the programme will be published elsewhere 
(Vanelle et al., in preparation).
In each field, there are now typically 10--20 calibration stars available
within the magnitude range $B=12$--20, with both $B$ and $V$ photometry
available.

With these sequences it was straightforward to compute photometric 
solutions for the $B_J$ isophotal magnitudes measured in the DSS 
direct plate scans. However, isophotal magnitudes of extended objects 
such as low-redshift QSOs with detectable host galaxy fuzz, but also
of single point sources with nearby projected neighbouring objects,
are systematically too bright.
Limiting the photometry to very small apertures was not acceptable
because of the small dynamic range and the coarse spatial resolution 
of the DSS. We therefore decided to seek for less biased photometry
of the survey targets.

Our adopted solution was to measure source fluxes in the digitised
spectra, by summing up pixel values over the $B_J$ band, and to
calibrate these magnitudes against the direct plates data.
This approach has several advantages:
\begin{itemize}
\item Since the spectral magnitudes are based on the `PSF
amplitude estimator' used for the extraction of spectra, 
they correspond to a photometric
aperture of the size of the seeing disk, thereby maximising the
contrast between nucleus and surrounding fuzz in low-$z$ QSOs and
Seyferts. 
\item Flux limits and photometry are defined in the same data sets,
minimising incompleteness near the limits due to photometric scatter.
\item Quasar selection and photometric measurements correspond to the
same epoch, entirely eliminating the effects of variability bias.
\item The dynamic range of the spectral plates is much larger than 
that of the (glass copy-based) DSS data and can be fully exploited 
not only for selection, but also for photometry.
\end{itemize}

We estimate the global photometric uncertainty of these measurements 
to be of the order of 0.2\,mag, including all the error sources.
\rbf{This is confirmed by comparing 400 stars with published photoelectric 
photometry from Kilkenny et al.\ (\cite{kilkenny*:97:ECS2}) and 
Preston et al.\ (\cite{preston*:91:HK}), which gives an overall 
rms scatter of 0.17~mag. Since this comparison involves over 50
different Schmidt fields, this scatter includes the contribution of
zeropoint errors. However,
} 
we cannot exclude at present that in a few 
($<10$) fields with exceptionally poorly sampled photometric sequences the
zeropoint uncertainties might be the dominant error sources; we plan to
obtain improved sequences for such fields in the future.

Based on this calibration, the definition of flux limits for each
survey field is straightforward.  
The distribution of actual limiting magnitudes for the 207 fields 
is shown in Fig.\ \ref{fig:maglim}.
The average magnitude limit is $B_J = 17.1$, approximately
$\sim 1$\,mag deeper than the PG survey.
Variable plate quality and seeing are the two effects chiefly 
responsible for the spread of flux limits.

To obtain unbiased magnitudes independent of the position in the sky,
we have estimated the foreground extinction $A_{B_J}$ 
from the measured column densities $N_H$ of Galactic neutral hydrogen, 
using the formula $A_{B_J} = 3.9 \times N_H / 58$
where $N_H$ is given in units of $10^{20}\,\mathrm{cm}^{-2}$ 
(cf.\ Bohlin et al.\ \cite{bohlin*:78:SIH}).
Table~\ref{tab:app:fields} lists $N_H$ and $A_{B_J}$ for each field.
The mean extinction value, averaged over all fields, is 0.23\,mag.

\begin{figure}[tb]
\epsfxsize=\hsize
\epsfclipon
\epsfbox[81 86 341 257]{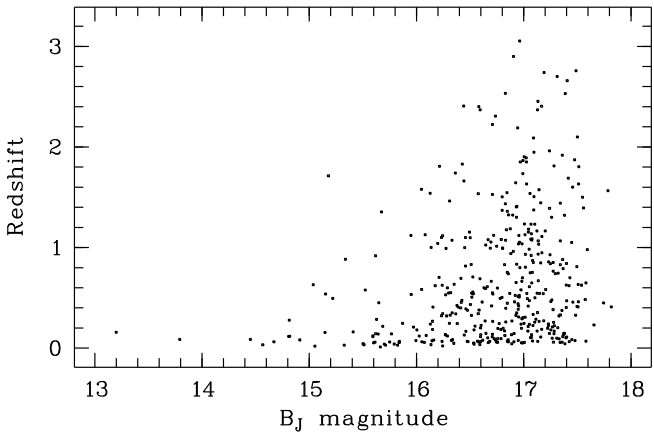} 
\caption[]{Hubble diagram of the 415 QSOs and Seyfert~1 galaxies 
forming the flux-limited sample. }
\label{fig:qhd}
\end{figure}

\section{The quasar sample   \label{sec:sample}}

Altogether, 415 QSOs and Seyfert~1 galaxies were identified in the surveyed
area with prism $B_J$ magnitudes brighter than the actual limits
given in Table \ref{tab:app:fields}.
No luminosity discrimination between QSOs and Seyferts was applied;
all objects displaying broad (FHWM $\ga 1000$ km/s) emission lines
in their follow-up spectra have been included, as well as all objects
classified either as QSO or as Seyfert~1 in the catalogue by 
V\'{e}ron-Cetty \& V\'{e}ron (\cite{vero+vero:96:AGN}).
The sample is listed in full in Table \ref{tab:app:qsos} of the
Appendix, giving positions, optical magnitudes, and redshifts.
A Hubble diagram is shown in Fig.\ \ref{fig:qhd}, 
the marginal distribution
of redshifts is given in Fig.\ \ref{fig:z}.

This is by far the largest compilation of optically bright quasars 
existing to date for which well-defined selection criteria and 
accurate flux limits have been formulated. It increases the earlier
HES sample published by K\"ohler et al.\ (\cite{koehler*:97:LQLF}) by
a factor of 6 in covered area and by a factor of 8 in sample size,
providing a much denser coverage of the Hubble
diagram particularly at low redshifts. Note that, although the fields
of K\"ohler et al.\ (\cite{koehler*:97:LQLF}) 
are all located within the present nominal survey
area, neither magnitudes nor effective areas are exactly identical.
The reasons for this are (i) the change from the $B$ band photometric
system to the $B_J$ passband, enforced by the change of direct plate
material, and (ii) the recent homogeneous rescanning and re-reduction of
all fields. The latter fact caused also a few of the objects of the 
K\"ohler et al.\ sample to be absent in the new sample (mainly because
of a slightly different `overlap' criterion), while some
others were added. These changes, however, are minor.

\section{Comparison with other surveys  \label{sec:others}}

There are not many surveys sensitive to the same domain in the sky
and in the magnitude/redshift plane as the HES. 
While this has allowed the HES to make 
many `discoveries', the possibilities to independently test selection 
efficiency and completeness are necessarily restricted.
We shall consider some of the available surveys in turn, and comment briefly 
on the balance of mutual detections.
Apart from the last subsection, we limit this comparison to surveys with
well-defined magnitude limits and selection criteria, i.e.\ we do not
consider visually selected samples as these are known to be highly
incomplete.

\subsection{The Palomar-Green survey}

With an average limiting magnitude of $B_{\mathrm{lim}}=16.16$ and an
effective area of more than 10\,000 deg$^2$
(Green et al.\ \cite{green*:86:PG}), the PG survey is
a cornerstone among quasar surveys. 
Of the $\sim 1800$ UV excess objects with stellar morphology, 
114 detected QSOs and Seyfert~1 galaxies form the 
`Bright Quasar Sample'
(BQS; Schmidt \& Green \cite{schm+gree:83:QEBQS}). 
There are some 30 PG fields in the region 
$9^{\mathrm{h}} < \alpha < 15^{\mathrm{h}}\,20^{\mathrm{m}}$,
$-12^\circ <\delta< +2\fdg 5$, corresponding to ESO fields
708--725, 780--797 and 850--869. The average PG limiting magnitude in
these fields is $16.14\pm 0.18$, but with an east-west trend; 
if bisected at $\alpha = 12^{\mathrm{h}}$, the limits become
$B_{\mathrm{lim}} = 16.31\pm 0.10$ (east), and
$B_{\mathrm{lim}} = 16.03\pm 0.15$ (west).
In these fields, the BQS contains 9 objects that have processable 
counterparts in the HES (i.e., that are not overlaps etc.),
all of which were recovered in the HES candidate selection.
However, the HES $B_J$ magnitudes clearly place 5 out of these 9 objects,
and possibly another one (depending on the $B-B_J$ correction),
\emph{below} the nominal PG limits. On the other hand, the HES has
identified 7 further QSOs and Seyferts that were missed in the PG.
Two of these are low-$z$ Seyferts which were probably excluded
by the morphological selection, yet the other five are clearly 
luminous QSOs. Only one of them is located close to the magnitude 
limit, for the others neither the uncertainties in the $B-B_J$ term 
nor variability can account for this effect.
It seems that the PG suffers severely from what appear to be random losses;
however, these are partly compensated, with respect to the number counts, 
by \emph{overcompleteness} from sources below the nominal survey limit.

\subsection{The Edinburgh Quasar Survey}

The EQS was designed as a multicolour survey over $\sim 330$ deg$^2$
close to the equator, covering ESO/SERC fields 789--794 and 861--867 with a
limiting magnitude $B<18.0$ (thus, full overlap of nominal area with
the HES). A bright subset of 12 QSOs with $B<16.5$ was presented
by Goldschmidt et al.\ (\cite{goldschmidt*:92:SDBQ}), 
\rbf{based on which
they inferred a very high degree of incompleteness for the BQS.
The HES quasar sample contains 8 QSOs out of the 12 in the Goldschmidt et
al.\ sample. One entry has to be deleted from 
this list: for Q\,1250$-$0700 we obtained a slit spectrum 
which shows an F-type star. 
The remaining three EQS objects are flagged as overlaps in the HES data.
The HES has found one additional QSO brighter than $B=16.5$ that is 
not contained in the EQS sample, probably because a prominent host galaxy 
led to its exclusion as non-stellar. Although the well-known 
blazar 3C\,279 is listed in the HES but not in Goldschmidt et al,
this object had in fact been detected by the EQS, but was not considered 
as proper member of the QSO class (Miller 2000, private communication).
A more detailed comparison of EQS and HES including also several unpublished 
QSOs showed that the photometric scales are in full agreement.
On the other hand, the extremely high concentration of bright QSOs 
found by Goldschmidt et al.\ (\cite{goldschmidt*:92:SDBQ})
could not be reproduced in the HES, mainly due to the fact that some
of the brightest low-$z$ QSOs appear to be systematically brighter 
in the EQS. The most likely explanation for these differences is
that it is caused by the different photometric techniques used
(isophotal in the EQS, point source matching in the HES).
}

\subsection{The `Large Bright Quasar Survey'}

The similarity of the HES selection strategy to that of the
LBQS is probably greater than to any other quasar survey
(cf.\ Hewett et al.\ \cite{hewett*:95:LBQS6}).
Despite its name, the LBQS has mainly yielded intermediate-brightness QSOs,
and its bright limit of $B_J > 16.0$ as well as the moderate survey area
of $\sim 450$\,deg$^2$ allows comparison between HES and the LBQS
only over a narrow range of magnitudes around $B_J\simeq 17$.
There are 7 fields (ESO field numbers 854, 858, 861, 863, 864, 867, 
and the SGP field) that both surveys have in common. 
All 10 LBQS quasars brighter than the HES completeness
limit were recovered and are part of the flux-limited sample.
The HES has identified further five relatively bright QSOs in these 
fields that are not part of the LBQS, of which two are at low redshift
($z<0.1$), and three have apparently been missed by the LBQS.
It is of course possible that some of these have been flagged as 
unprocessable in the LBQS reduction.

\subsection{The quasar catalogue of V\'{e}ron-Cetty \& V\'{e}ron}

As a routine task during the HES data reduction, a general 
comparison with the latest version of the 
catalogue by V\'{e}ron-Cetty \& V\'{e}ron
(\cite{vero+vero:96:AGN}) has always been performed. The catalogue
was cross-correlated with the input source list from the 
direct plate object search, followed by a forced extraction
of all objects within a large error box of $30''\times 30''$
around the nominal AGN position -- thus avoiding to make any 
prior assumptions about the integrity of the corresponding spectra.
The extracted spectra were inspected, and those contaminated
by overlapping neighbour objects were eliminated.
Altogether, 143 sources with processable spectra above
the `completeness limits' were identified; of these, 139
had been included into the QSO candidate lists. Of the 
remaining four, three had actually been selected but were 
\rbf{missed} in the visual inspection phase, and only one object
was missed by the selection criteria. This last object is
a bright $z=3.3$ quasar with a very red and short continuum,
but a prominent Ly$\alpha$ line that happens to lie just 
outside the wavelength region used for the emission line
detection, indicating the approximate high-redshift limit
for our selection criteria.

\section{Redshift-dependent selection efficiency}

Optically selected quasar samples are often under
suspicion of substantial incompleteness and
selection biases varying with redshift, because of
the difficulty of discriminating quasar candidates against 
the outnumbering majority of normal stars. In this section we
investigate the importance of such biases for the HES,
and discuss the limitations of our adopted QSO survey scheme
in this context.

\begin{figure}[tb]
\epsfxsize=\hsize
\epsfclipon
\epsfbox[70 87 341 258]{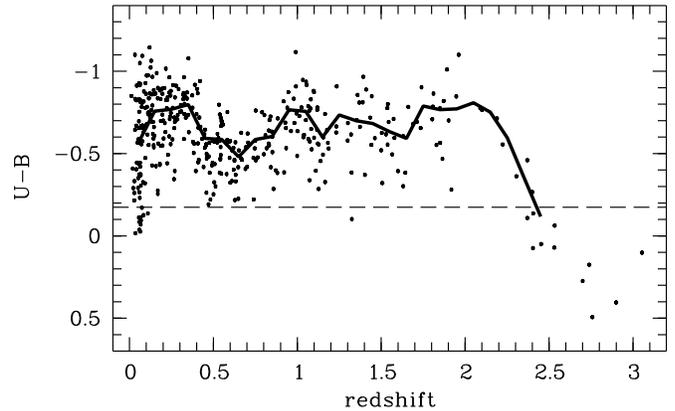} 
\caption[]{Distribution of UV excess ($U-B$) as a function of redshift.
The thick solid polygon traces the distribution mean in bins of 
$\Delta z = 0.1$; the horizontal dashed line marks the adopted 
UVX selection criterion of the HES.}
\label{fig:qsocolours}
\end{figure}

Fig.\ \ref{fig:qsocolours} shows the variation of UV excess with
redshift, for the 415 objects in the HES flux-limited sample. 
This plot has been prepared by converting the excess over the
UV selection criterion into a standard ($U-B$) colour, using the
empirical relation from Fig.\ \ref{fig:uminbcal}.
The diagram shows that the mean ($U-B$) varies systematically 
over the sampled redshift range, but stays always well above the
HES selection limit. However, some objects have a very weak UV excess,
close to or even below the relatively relaxed selection limit of the HES. 
These objects entered the sample by matching at least one of the further 
selection criteria employed by the HES. This is particularly relevant for 
two groups of objects: 
\begin{itemize}
\item At low redshifts, Seyfert galaxies tend
to lose their UVX as the nuclei get fainter relatively to their 
host galaxies. To a certain extent, this effect has been counteracted 
by our dedicated Seyfert criterion `quartile distance' 
(cf.\ Sect.\ \ref{sec:coloursel}).
\item At the highest redshifts,
$z\ga 2.4$, UVX selection of QSOs is known to be impossible.
For $z\simeq 2.5$, the redder \texttt{hpp2} criterion is still
effective; for $z>2.6$, the objects are selected only by the 
feature detection algorithm. 
\end{itemize}
Apart from these two extremes, the distribution of 
measured ($U-B$) values around the mean relation is nearly Gaussian with
$\sigma_{U-B}=0.15$~mag, and the selection boundary is always separated
by more than $\sim 2\sigma_{U-B}$. This is qualitatively confirmed 
by Fig.\ \ref{fig:qsocolours}, where at no redshift the distribution
appears to be truncated by the selection limit. While the mere 
presence of some objects with colours close to critical 
give rise to the suspicion that there may be others that have just 
been missed, there is no evidence that such objects could make up
a sizeable fraction of the population. Note, however, that a more
restrictive selection criterion such as $(U-B)<-0.4$ conventionally 
applied in UVX surveys would have caused severe incompleteness in
certain redshift regions.

\begin{figure}[tb]
\epsfxsize=\hsize
\epsfclipon
\epsfbox[72 86 341 257]{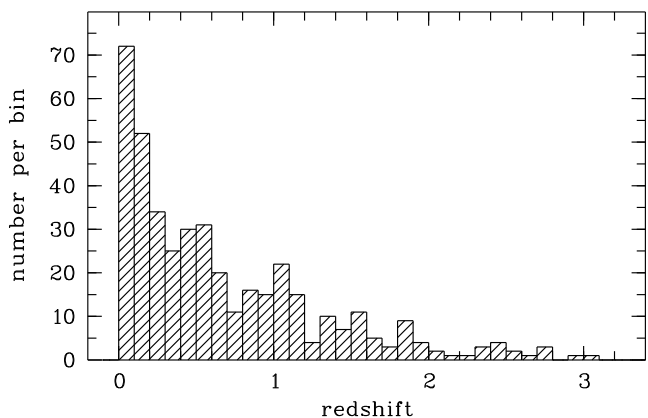} 
\caption[]{Redshift distribution of the HES flux-limited sample.}
\label{fig:z}
\end{figure}

A further simple and very common test for redshift-dependent 
selection biases is to look for excess or depletion bins 
in the observed redshift histogram.
The distribution displayed in Fig.\ 
\ref{fig:z} shows a steady decline in numbers from low to
high redshifts, without indication of significantly high or low
bins. The deviations from a smooth `fit by eye' are entirely
consistent with Poissonian ($\sqrt{n}$) shot noise, and
there is no evidence for a strong $z$ dependence of
the survey selection function. 

Notorious for UVX surveys is the regime around $z\approx 0.7$. 
While the number of HES QSOs in this redshift bin indeed shows a minimum, 
its value with respect to the neighbouring bins is not significantly low. 
It cannot be excluded that the selection efficiency for quasars in this 
redshift region may be somewhat reduced, as the mean colours get slightly
redder (Fig.\ \ref{fig:qsocolours}). 
However, similar features of comparable formal significance 
occur in other redshift bins that do not correspond to 
identifiable features in the colour distribution and are
almost certainly due to statistical fluctuations.
This is underlined by the distribution of apparent magnitudes vs.\
redshifts in Fig.\ \ref{fig:qhd}, where some of the `gaps' 
and `excesses', including those at $z\approx 0.7$ and $z\approx 1$, 
occur mainly among the \emph{brighter} objects, while any selection biases 
would be expected to be stronger close to the limiting magnitude 
where the S/N is low. Again, this does not necessarily imply a selection 
function of unity,  but it supports our contention that redshift-dependent 
selection  effects are not important for this sample 
except near the high-$z$ limit.

The situation for $z>2$ requires a separate discussion. Apart from
the colour changes due to a simple shift of the spectrum,
real spectra are also affected by intergalactic Lyman forest absorption.
Our set of colours is capable, in principle, of selecting
QSOs up to redshifts as high as 2.8, 
but this is too optimistic for real spectra. 
Nevertheless, \emph{all} quasars with $z<2.6$ in the current sample 
are strictly colour-selected, although several have also emission-line 
detections. In contrast, all $z>2.6$ objects were detected 
only by their broad Ly$\alpha$ emission and the corresponding break 
feature on its blue side.
Because of the complicated dependency of the feature detection
threshold as a function of wavelength 
(cf.\ Sect.\ \ref{sec:cansel}), 
the true number of high-$z$ QSOs within the flux limits is
necessarily underestimated in the present sample.
The degree of this incompleteness depends on redshift, 
signal-to-noise ratio, but also on the intrinsic properties of 
the QSOs, the equivalent widths of their Ly$\alpha$ emission lines, 
and the strength of the Lyman forest break features. 
Only a detailed simulation study over this multiparameter space 
will provide reliable estimates of the degree of incompleteness, 
and generate accurate surface and space densities of quasars with 
redshifts $\ga 3$.

\section{The surface density of bright QSOs \label{sec:surfdens} }

\begin{figure}[tb]
\epsfxsize=\hsize
\epsfclipon
\epsfbox[76 85 341 257]{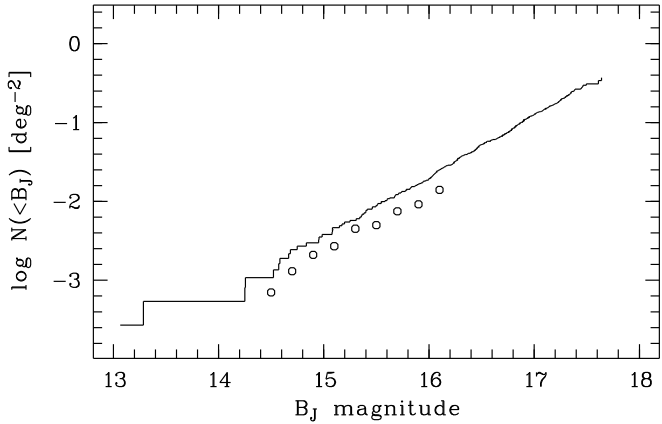} 
\caption[]{Cumulative surface density $N(<B_J)$ 
of bright QSOs as a function of
magnitude, without cut in redshift or luminosity. 
The small circles show the corresponding relation from Schmidt \& Green 
(\cite{schm+gree:83:QEBQS}), with magnitudes corrected as discussed
in the text.}
\label{fig:surfdens1}
\end{figure}

The empirical relation between fluxes and
number of sources per unit solid angle is
an important diagnostic tool for QSO samples. Once field
properties and sample photometry are complete, the
cumulative surface density of QSOs brighter than magnitude $B_J$,
$N(<B_J)$, can be easily computed from summing over 
$1/\Omega_{\mathrm{eff}}$ for all relevant sample objects.
We have derived the $N(<B_J)$ relation from the 
`complete sample' presented above, shown by the solid line in
Fig.~\ref{fig:surfdens1}. The abscissa values
in this relation are extinction-corrected magnitudes.
The bright end of the curve is dominated
by the extremely luminous sources 3C~273 and HE~1029$-$1401, 
the two brightest QSOs known in the sky, but the main part 
can be well approximated by a simple power law of slope 
$\beta = 0.77\pm 0.01$ ($\log N(<B_J) = \beta B_J + \mathrm{const}$).
This slope is slightly steeper than the extremely flat relation 
($\beta = 0.67$) that we had obtained in an earlier analysis 
(K\"ohler et al.\ \cite{koehler*:97:LQLF}). 
The surface densities around $B_J\la 15$ are now somewhat lower, 
while those for $B_J\la 17$ are even slightly higher than measured 
before. However, the differences are within the formal uncertainties.
The result for the zeropoint is $\log N(<16) = -1.68$, i.e.\
we find one QSO brighter than $B_J = 16$ per 48\,deg$^2$.

Comparison relations from other surveys are numerous for fainter 
magnitudes, but rare for $B_J\la 16.5$. Additional complications
arise because various conventions are used for the
treatment of low-luminosity and/or low-redshift AGN in the samples:
Some authors apply prior cuts in absolute magnitude -- e.g., 
Hartwick \& Schade (\cite{hart+scha:90:QE}) used only objects with
$M_B<-23$ in their compilation. Some others do not count low-redshift
AGN, irrespective of absolute magnitude. All these filters
are particularly effective at bright apparent magnitudes.

The Palomar-Green BQS (Schmidt \& Green \cite{schm+gree:83:QEBQS})
provides the only available optically selected QSO sample 
with a well-defined flux limit subtending over a similar region 
in the Hubble diagram. Their published cumulative surface densities
include low-redshift and low-luminosity AGN and can be compared
to the HES counts of Fig.\ \ref{fig:surfdens1} (the BQS is limited 
to $z<2.2$, but this filter has negligible effect at bright magnitudes).
However, for a proper comparison the magnitudes of Schmidt \& Green 
need to be corrected for Galactic extinction.
We have estimated the mean offset to $\Delta B \simeq 0.2$ 
from comparing the uncorrected Schmidt \& Green counts
with the extinction-corrected surface densities listed by 
Hartwick \& Schade (\cite{hart+scha:90:QE}),
which for $B < 16$ are entirely dominated by the BQS. 
The BQS magnitudes were shifted by another 0.1\,mag
in order to approximately correct for the different
photometric systems ($B_J$ vs.\ $B$; 
\rbf{cf.\ Hewett et al.\ \cite{hewett*:95:LBQS6}}), 
so that the resulting number-magnitude relation for the BQS, 
shown by the small circles in Fig.~\ref{fig:surfdens1},
is based on an adopted magnitude scale 
$B_J \simeq B_{\mathrm{BQS}}-0.3$.

\begin{figure}[tb]
\epsfxsize=\hsize
\epsfclipon
\epsfbox[72 86 341 257]{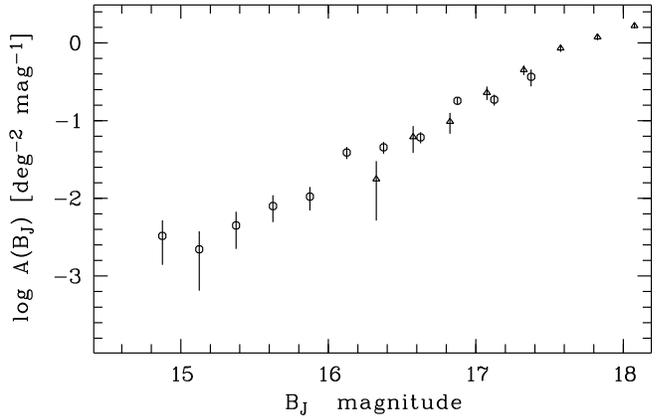} 
\caption[]{Differential surface densities of QSOs for the
redshift interval $0.2<z<2.2$, with error bars from Poisson
statistics. Circles denote the HES measurements, triangles show 
the corresponding data from the LBQS 
(Hewett et al.\ \cite{hewett*:95:LBQS6}).}
\label{fig:surfdens2}
\end{figure}

The surface densities of the BQS are below those of the HES
at all covered magnitudes, whereas the inferred slopes are virtually
identical. Therefore the ratio of surface densities is nearly
constant, $N_{\mathrm{HES}}/N_{\mathrm{BQS}} = 1.48\pm 0.06$.
While this excess is highly significant in itself, it is much lower
than the earlier claimed incompleteness rate of a factor 3 or more 
by Goldschmidt et al.\ (\cite{goldschmidt*:92:SDBQ}) or by ourselves
(K\"ohler et al.\ \cite{koehler*:97:LQLF}). As discussed above,
the small EQS sample used by 
Goldschmidt et al.\ (\cite{goldschmidt*:92:SDBQ}) is basically a subset of
the new HES sample, and their high number seems to constitute a statistical
fluctuation. Neither Goldschmidt et al.\ nor K\"ohler et al.\
included Galactic extinction corrections. For the K\"ohler et al.\ sample,
this would have reduced the inferred incompleteness to a factor $\sim 2.4\pm 0.5$,
statistically consistent with the new value of 1.5 on the 2$\sigma$ level.
\rbf{
It is possible that incompleteness in the BQS 
is higher for low-luminosity sources, but the direct object-to-object
comparison in the overlapping area does not suggest a strong effect.
For the present paper we wish to limit this discussion to an 
overall comparison without splitting samples according to nuclear luminosities,
which is problematic without going into the details of
comparing our point-source with the BQS isophotal magnitudes.
This way we also avoid the complications of a model-dependent 
conversion to absolute magnitudes. 
We are currently investigating the host galaxy properties of HES and
BQS selected QSOs, which will allow us to understand
the possible selection effects in much greater detail.
}

These new results show with high statistical significance 
that the BQS is incomplete, but also
that the degree of incompleteness has recently been overestimated
\rbf{(a similar conclusion was reached by Mickaelian
et al.\ \cite{mickaelian*:99}).
}
On the other hand, our cross-comparison between
HES and BQS in overlapping fields has yielded a significant 
contamination of the BQS with objects fainter than the nominal limits.
It thus appears that there is also evidence for
\emph{overcompleteness} in the BQS, to a much higher degree than 
the 15\,\% estimated by Schmidt \& Green (\cite{schm+gree:83:QEBQS}).
The HES--BQS comparison suggests
that up to $\sim 50$\,\% of the BQS quasars may be affected; 
in that case, the \emph{true incompleteness} of the BQS
would be again at least by a factor of $\ga 2$.
To assess the origin of this overcompleteness is difficult.
One possibility was suggested by K\"ohler et al.\ (\cite{koehler*:97:LQLF}) 
in order to explain the different luminosity function slopes, 
invoking subtle photometric biases favouring moderately extended
low-redshift QSOs over real point sources. Alternatively or additionally,
the photometric errors in the PG survey might simply be much larger
than assumed by Schmidt \& Green (\cite{schm+gree:83:QEBQS}).

At fainter magnitudes, the HES counts are in excellent agreement 
with most other surveys. As an example, in Fig.\ \ref{fig:surfdens2}
we compare the HES \emph{differential} number counts for the redshift range 
$0.2<z<2.2$ to the same relation for the LBQS published by
Hewett et al.\ (\cite{hewett*:95:LBQS6}). Since the LBQS magnitudes
are also given in the $B_J$ system, no photometric transformation is 
required; a small shift of the LBQS values, by $\sim 0.1$\,mag
towards brighter magnitudes, has been applied to correct 
for Galactic extinction neglected by Hewett et al.
The derived surface densities are statistically consistent
in all magnitude bins. The brightest LBQS value seems to be somewhat low,
possibly indicating incompleteness caused by saturation effects
during plate digitisation with the APM machine as discussed by Hewett
et al., but even that data point is not formally inconsistent with 
the HES.

\section{Conclusions}

The Hamburg/ESO survey is an ongoing project that provides substantial 
improvements over previous searches for optically bright QSOs.
We have constructed a new flux-limited quasar sample useful
for a wide variety of statistical studies, consisting of 415 objects 
distributed over 3700~deg$^2$ in the sky. Greatest care was taken to
minimise selection biases that could lead to redshift-dependent 
incompleteness. Key properties of the HES in this respect are 
(i) very relaxed UV excess selection criteria, equivalent to $(U-B)<-0.18$, 
(ii) the small errors of determining the UV excess in individual objects,
and (iii) the high spectral resolution of the ESO objective prism spectra 
which ensures low stellar contamination without sacrifices in completeness.

Comparing the sample with the results of other surveys, in terms
of general number counts as well as by detailed comparison of common
survey regions, we find no evidence that the HES might miss a sizeable 
fraction of the known QSO population, within its redshift limits.
The surface density of detected bright quasars, for given magnitudes, 
is consistent with most recently published work, but the HES sample
greatly improves the statistics at very bright magnitudes.
We confirm that the Palomar-Green BQS is incomplete (although the
degree of incompleteness has been overestimated in the past), 
but alert the reader that there is also evidence for significant 
overcompleteness.

Several precautions were taken to reduce potential selection biases
associated with low-redshift QSOs.
Unlike most other optical QSO surveys, there is no discrimination of 
objects with extended morphological structure; these objects are subjected
to the same selection criteria as the point sources.
In the extraction of spectra, nuclear properties are enhanced
relative to possibly `red' galaxy contributions. This also affects
the derived magnitudes which are dominated by the nuclear emission.
Furthermore, a specific selection criterion has been implemented that is 
sensitive to `Seyfert-like' spectral shapes. All these factors enable the 
HES to produce the first well-defined optically selected
samples of low-$z$ QSOs and bright Seyfert~1 nuclei. At the same time,
selection biases against gravitationally lensed QSOs are avoided.

The HES substantially improves the sampling in the high-luminosity, 
high-redshift domain unaccessible to traditional UV excess surveys. 
While the number of $z>2.5$ sources in the present flux-limited sample 
is still small, there is a large reservoir of unambiguously detected 
high-$z$ quasars. By using dedicated flux limits optimised for these objects, 
combined with a detailed determination of the survey selection function, 
the total number of bright QSOs in the redshift range $2.5\la z\la 3.3$ 
that are part of a well-defined flux-limited sample will
be increased by at least an order of magnitude in the near future.

\acknowledgements{Thanks to Drs.\ T.~Beers and D.~Kilkenny 
who made the results of their photoelectric photometry available 
in digital form, 
\rbf{and to Dr.\ L.~Miller for allowing us to compare
results with unpublished data from the Edinburgh Quasar Survey. 
}
We are indebted to Dr.~H.-J. Hagen who developed 
and maintained the operating software to perform the full-matrix 
digitisation of Schmidt plates with the Hamburg PDS.
The Hamburg/ESO survey would not have been possible without 
the continued support by ESO staff members, in particular 
H.-E. Schuster, B. Reipurth, and O. and G. Pizarro 
who took all the Schmidt plates. Substantial observing time
was allotted to this project as ESO key programme 02-009-45K.
Parts of this work were supported by the DFG under grants Re 353/33
and Re 353/40. We also acknowledge support from BMBF/DLR grants
05-5HH41A and 50\,OR\,9606\,O.
}

\newpage

\appendix

\section{HES survey fields}

Table \ref{tab:app:fields} lists all 207 Hamburg/ESO survey fields 
used in this investigation. The first two columns list
the field number in the standard ESO/SERC numbering system, and the 
ESO plate number of the objective prism plate. Next given are right
ascension and declination of the nominal field centres, in B1950
coordinates. Note that the actual plate centres may deviate from these
nominal positions by several arc minutes. The following column gives the 
limiting magnitudes per field (`completeness limit' = 5$\sigma$ detections) 
as obtained from the photometric calibration. The last two entries feature
the column density of neutral hydrogen in $10^{20}\,\mathrm{cm}^{-2}$
and the adopted correction for Galactic extinction in the $B_J$ band.

\begin{table}[htb]
\caption{\rbf{Table is only available in electronic form at the CDS
via anonymous ftp to \texttt{cdsarc.u-strasbg.fr}
}}
\label{tab:app:fields}
\end{table}

\section{The QSO sample}

Table \ref{tab:app:qsos} presents the flux-limited sample of 415 QSOs
and bright Seyfert~1 galaxies selected from the Hamburg/ESO survey. 
The designation is listed in the first two columns: first the standard
HE + B1950 coordinate string used by the HES, followed by the name
as it appears in the catalogue of V\'{e}ron-Cetty \& V\'{e}ron 
(\cite{vero+vero:96:AGN}). The next columns give J2000 equatorial coordinates
obtained from the DSS plate solutions, and 
the ESO field number where the QSO was found. \rbf{The following two columns  
list the redshifts and a flag indicating if the redshift was derived from
our own slit spectroscopy; otherwise, the $z$ value from 
V\'{e}ron-Cetty \& V\'{e}ron (\cite{vero+vero:96:AGN})
was used. The final two columns contain the small-aperture
} 
$B_J$ apparent magnitudes, and extinction-corrected $B_J$ values.

\begin{table}[htb]
\caption{\rbf{Table is only available in electronic form at the CDS
via anonymous ftp to \texttt{cdsarc.u-strasbg.fr}
}}
\label{tab:app:qsos}
\end{table}

\end{document}